\documentclass[twocolumn]
{aastex631}
\usepackage{amsmath}
\usepackage{amssymb}
\usepackage{aas_macros}
\usepackage{appendix}
\usepackage{gensymb}

\begin{document}

\title{A Multi-messenger Search for a Nearby Microquasar Contributor to the Cosmic Ray Knee}

\correspondingauthor{Si-Ming Liu}
\email{liusm@swjtu.edu.cn}
\correspondingauthor{Yi-Qing Guo}
\email{guoyq@ihep.ac.cn}
\correspondingauthor{Hua Yue}
\email{yuehua@ihep.ac.cn}

\author{Lin Nie}
\affiliation{School of Physical Science and Technology, Southwest Jiaotong University, Chengdu, 610031, China}

\affiliation{State Key Laboratory of Particle Astrophysics, Institute of High Energy Physics, Chinese Academy of Sciences, Beijing, 100049, China
}
\author{Hua Yue}
\affiliation{State Key Laboratory of Particle Astrophysics, Institute of High Energy Physics, Chinese Academy of Sciences, Beijing, 100049, China
}
\affiliation{University of Chinese Academy of Sciences, Beijing 100049, China}

\author{Yi-Qing Guo}
\affiliation{State Key Laboratory of Particle Astrophysics, Institute of High Energy Physics, Chinese Academy of Sciences, Beijing, 100049, China
}
\affiliation{University of Chinese Academy of Sciences, Beijing 100049, China}
\affiliation{TIANFU Cosmic Ray Research Center, Chengdu 610000, China}
\author{Si-Ming Liu}
\affiliation{School of Physical Science and Technology, Southwest Jiaotong University, Chengdu, 610031, China}

\begin{abstract}
Recently, LHAASO has detected five microquasars with high confidence, which are associated with SS 433, V4641 Sgr, GRS 1915+105, MAXI J1820+070, and Cygnus X-1, respectively. Except for Cygnus X-1, the maximum energies of gamma-ray photons emitted from these sources all exceed 100 TeV, strongly suggesting that microquasars are capable of accelerating cosmic-ray particles to energies above the PeV range. This work investigates the origin of the cosmic-ray knee region based on gamma-ray observational data from the aforementioned sources, combined with cosmic-ray proton, helium, and all-particle energy spectra, as well as anisotropy observations. Calculations indicate that these known sources contribute negligibly to the cosmic-ray knee region. However, further joint analysis reveals that a single microquasar located in a region approximately on the 2.6 kiloparsec scale in the anti-Galactic center direction can reasonably reproduce the observed cosmic-ray proton, helium, and all-particle energy spectra, as well as anisotropy features detected near Earth. We propose that this region may host one or several unidentified microquasars or similar systems, whose accelerated cosmic rays could dominate the observational characteristics of the knee region.

\end{abstract}


\section{Introduction} \label{sec:intro}
One of the most prominent features in the cosmic-ray energy spectrum is the “knee”-like structure observed around ~ $\rm PeV$ energies, known as the cosmic-ray knee. In this energy range, the spectral index of the cosmic-ray spectrum steepens dramatically from about $-2.7$ below $\sim 4~\rm PeV$ to about $-3.1$ above $\sim 4~\rm PeV$ \citep{2024PhRvL.132m1002C,1959bSov.Phys.JETP...35..441}. The exact physical origin of the knee structure remains unclear, and various explanations such as cosmic-ray acceleration origins and propagation origins have been proposed. Although recent studies tend to argue against propagation origins \citep{2026arXiv260117851Y}, no unified theoretical framework has been established.

From an observational perspective, although cosmic rays lose directional information about their sources during propagation through interstellar magnetic fields, very-high-energy gamma rays—produced as secondary radiation during cosmic-ray acceleration and propagation—can directly indicate the presence of PeVatrons in local regions. Therefore, observing gamma rays above $100~\rm TeV$ using high-sensitivity detectors is an effective approach to investigating the origin of the knee.

The Large High Altitude Air Shower Observatory (LHAASO) is a gamma-ray detection facility covering energies from $~\rm TeV$ to the PeV range, with an effective instrumented area of about one square kilometer. It achieves a point-source sensitivity of $\rm 10^{-14}~erg~cm^{-2}~s^{-1}$ above 100 TeV \citep{Cao_2025}, making it an ideal instrument for detecting ultra-high-energy gamma-ray sources. The observational data it has already accumulated provide crucial support for studying PeV cosmic rays and the origin of the knee structure.

Traditionally, supernova remnants (SNRs) have been regarded as the primary accelerators of cosmic rays in the Galaxy \citep{2010ApJ...718...31P,2012APh....39...12Z,2023ApJ...952..100N}. This view is largely based on the possibility that nonlinear diffusive shock acceleration could accelerate particles to PeV energies and explain the multi-wavelength non-thermal radiation from SNRs. Furthermore, the “standard propagation model,” which treats SNRs as the dominant sources, is consistent to some extent with observations of cosmic rays and diffuse Galactic gamma rays. However, recent studies indicate that only a very small fraction of supernovae can accelerate cosmic rays to knee energies or beyond \citep{2013MNRAS.431..415B,2018MNRAS.479.4470M}. For instance, only core-collapse supernovae occurring in dense stellar wind environments may reach or exceed PeV energies, while most supernovae can only accelerate particles to several hundred TeV, about an order of magnitude below the knee energy.

At the same time, recent LHAASO observations have shown that out of 12 reported microquasars \citep{Cao_2025}, five exhibit radiation above 100 TeV. In particular, the extended emission from the central region of SS 433 is spatially correlated with local gas clouds, suggesting a hadronic origin. This indicates that microquasars could be significant contributors to cosmic rays at PeV or even higher energies. Notably, LHAASO data reveal that the average logarithmic mass of cosmic rays decreases between 300 TeV and 3 PeV \citep{2024PhRvL.132m1002C,2024JHEAp..44..116L}, while the all-particle spectral index remains almost unchanged over the same energy range, hinting at a possible new spectral component near the knee.

Additionally, studies have found that unlike the smoother distribution at GeV energies, Galactic cosmic rays in the knee region and above exhibit pronounced inhomogeneity and clumpiness \citep{2023arXiv230510251G}. The PeV cosmic-ray flux observed at Earth may originate predominantly from a single nearby source. Therefore, it is necessary to search for such a source consistent with experimental observations.

In this work, we assume that cosmic rays in the sub-PeV range are mainly contributed by SNRs, with their spectra cutting off below PeV energies, while cosmic rays at PeV and above in the Galaxy are dominated by a single microquasar (or a similar binary system) relatively close to Earth. Based on LHAASO observations of ultra-high-energy gamma rays from microquasars, as well as data from other experiments on cosmic-ray proton, helium, all-particle energy spectra, and anisotropy, we evaluate the contribution of individual known microquasars emitting above 100 TeV to the cosmic-ray knee region and explore the microquasar origin of the knee spectral structure.

The paper is organized as follows: Section \ref{sec:method} introduces the methodology and framework; Section \ref{sec:result} presents and discusses the results; Section \ref{sec:conclusion} provides a summary and outlook for future work.

\begin{table}[htbp]
\footnotesize
\centering
\caption{Parameters of the propagation model.}
\label{tab1}
\tabcolsep 3.5pt
\begin{tabular*}{0.48\textwidth}{lcccccc}    
\hline \hline
$D_0{ }\left[\mathrm{cm}^{2} \mathrm{~s}^{-1}\right]$ & $\delta$ & $\nu_1$ & $\nu_2$ & $\mathrm{R_{br}}[\mathrm{GV}]$ & $v_A\left[\mathrm{~km} \mathrm{~s}^{-1}\right]$ & $R_{cut}[\mathrm{TV}]$ \\
\hline $3.2 \times 10^{28}$ & 0.33 & 2.48 & 2.38 & 335 & 25 & 200.0 \\
\hline 
\end{tabular*}
\end{table}

\begin{figure}[t]
    \centering
    \includegraphics[width=0.95\linewidth]{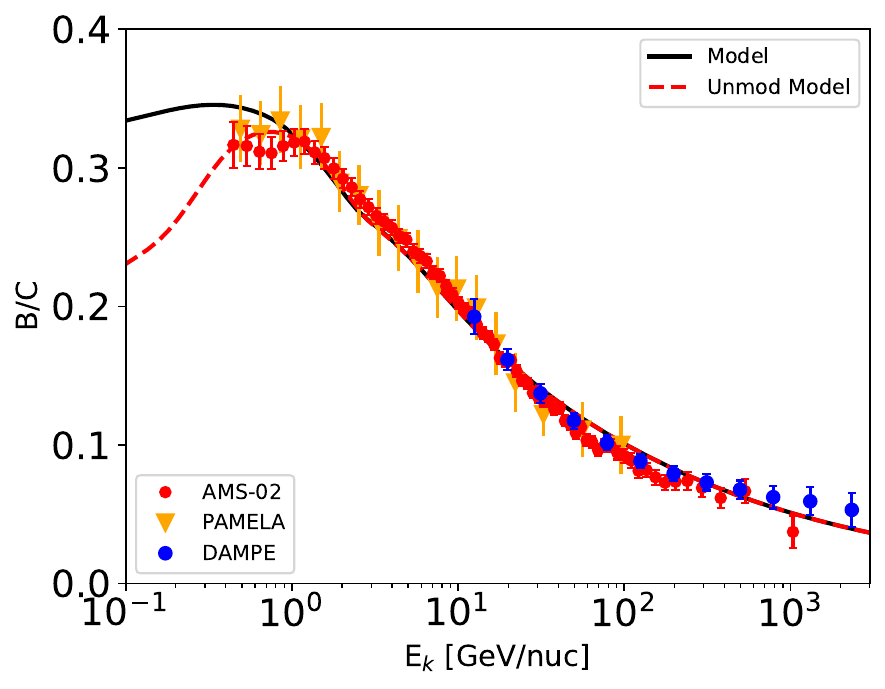}
    \caption{A comparison of the B/C ratio calculated using the CR propagation model with observational data from AMS-02 \citep{2017PhRvL.119y1101A}, PAMELA \citep{2014ApJ...791...93A}, and DAMPE \citep{2022SciBu..67.2162D}. The red dashed line indicates the spectrum calculated without considering solar modulation. In this study, the solar modulation potential is consistently assumed to be 550 MeV. }
    \label{BC}
    \end{figure}
\begin{figure*}[t]
    \centering
    \includegraphics[width=0.95\linewidth]{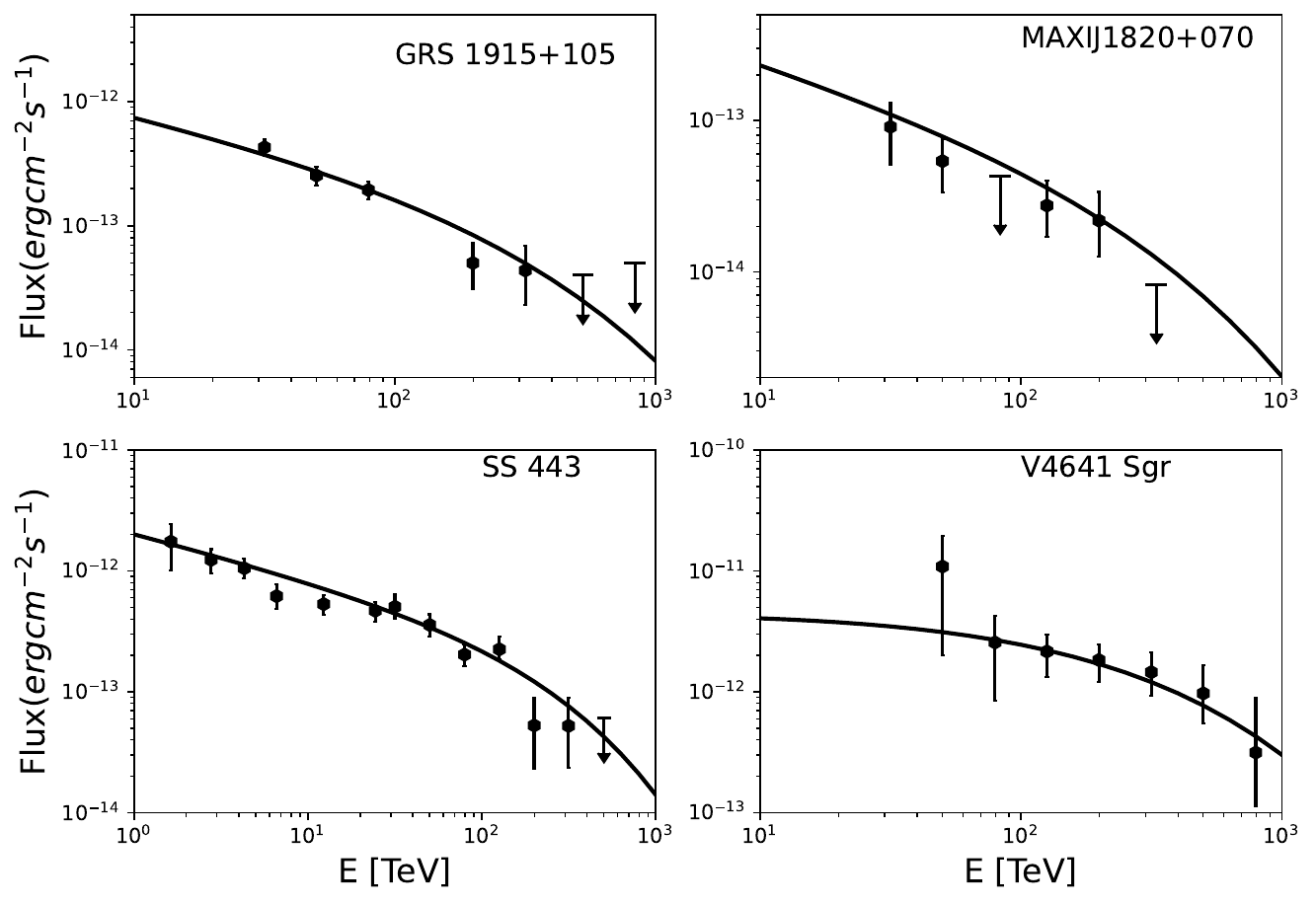}
    \caption{ The $\gamma$-ray of four microquasars predicted by the model compared with data observed by the LHAASO \citep{Cao_2025}.}
    \label{gamma}
    \end{figure*}

\begin{figure}[t]
    \centering
    \includegraphics[width=0.95\linewidth]{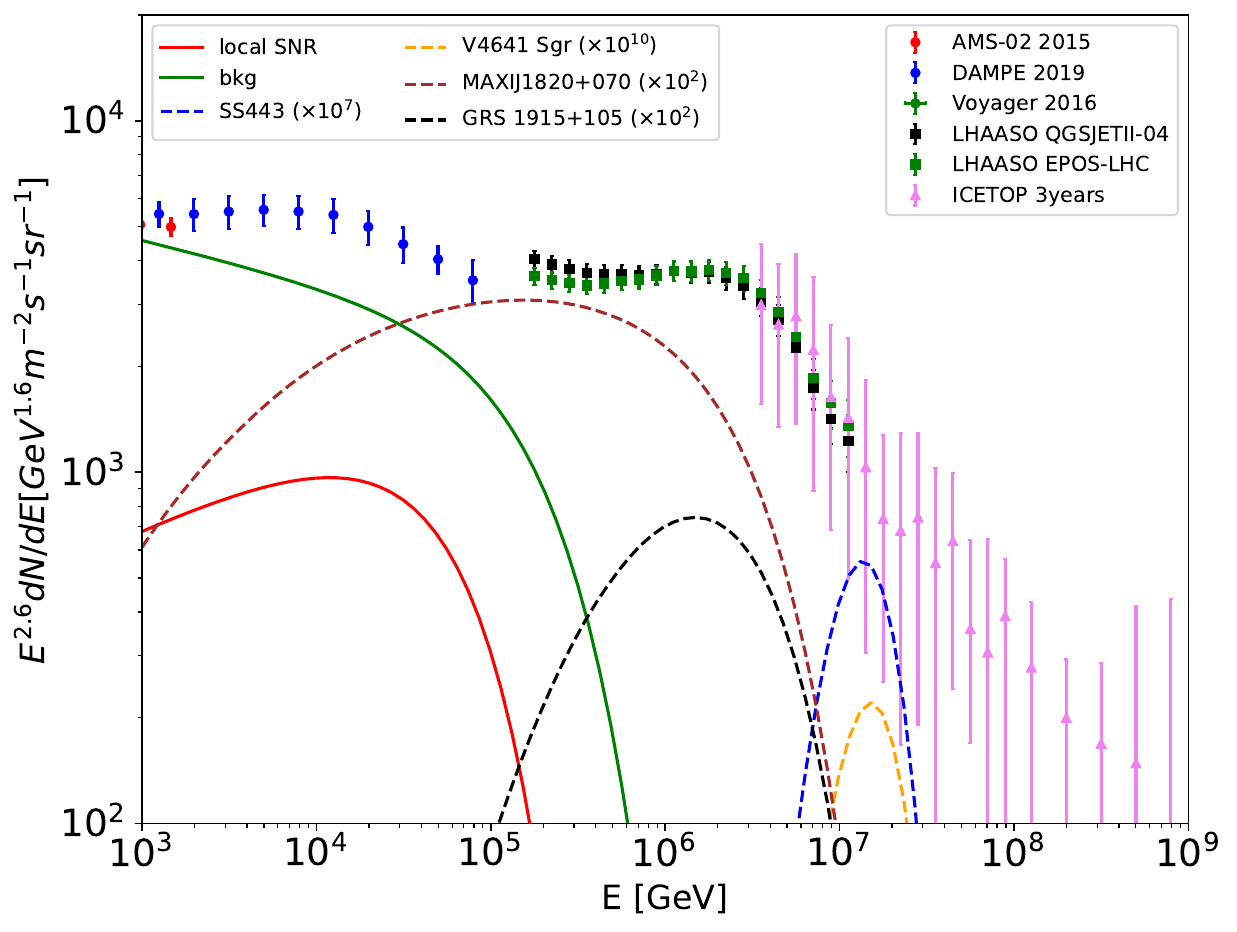}
    \caption{The CR proton spectrum is predicted through the CR propagation model and compared with observational data from AMS-02 \citep{2015PhRvL.114q1103A}, DAMPE \citep{2019SciA....5.3793A}, LHAASO \citep{Cao_2025} and IceTop \citep{2019PhRvD.100h2002A}. The green and red solid lines present background CR component and distribution from the local SNRs, respectively. The dashed lines show the distribution from the SS 433 (blue), V4641 Sgr (yellow), GRS 1915+105 (black), and MAXI J1820+070 (brown), respectively.}
    \label{proton1}
    \end{figure}

\section{MODEL \& METHODOLOGY} \label{sec:method}
We propose that the Galactic cosmic-ray flux observed at Earth is primarily composed of two components. Consistent with the standard cosmic-ray propagation model, cosmic rays below the knee region are mainly accelerated and dominated by SNRs, which we refer to in this work as the background component. The Galactic cosmic rays above the knee are contributed solely by one or a few isolated microquasars or microquasar-like binary systems located near Earth.

Based on this framework, this study first utilizes recent gamma-ray observational data from LHAASO for SS 433, V4641 Sgr, GRS 1915+105, and MAXI J1820+070 to individually constrain the cosmic-ray spectra accelerated by each source. Then, within the same propagation environment undergone by the background cosmic rays, the contribution of each source to the cosmic-ray flux at Earth is calculated. Finally, to simultaneously account for the observed cosmic-ray proton, helium, and all-particle spectra, as well as the amplitude and phase of the anisotropy, we assume the existence of a single, yet unidentified microquasar or similar source and use the observed anisotropy amplitude and phase as constraints to infer its possible spatial location.
\begin{table}[htbp]
\footnotesize
\centering
\caption{The luminosity and proton spectral parameters employed for the different sources are shown in Fig. \ref{gamma}.}
\label{tab2}
\tabcolsep 3.5pt
\setlength{\tabcolsep}{18pt}
\begin{tabular*}{0.45\textwidth}{lccc}    
\hline \hline
name & $\beta$ & $L_0$ $[ergs^{-1}]$   \\
\hline GRS 1915+105 & 2.5 & $1.5\times10^{39}$  \\
\hline MAXIJ J1820+070 & 2.5 & $1\times10^{38}$  \\
\hline SS 443 & 2.5 & $2\times10^{39}$  \\
\hline V4641 Sgr & 2.2 & $5\times10^{38}$  \\
\hline 
\end{tabular*}
\end{table}

\subsection{Propagation Model of  Background CRs}
Cosmic rays originate from astrophysical sources such as supernova remnants \citep{2023ApJ...952..100N,2012APh....39...12Z}, strong stellar winds from massive stars \citep{2019NatAs...3..561A},  rapidly rotating pulsars \citep{2021Natur.594...33C,2022ApJ...924...42N} and active microquasar jets \citep{Cao_2025,2026ApJ...997..163Y}. These energetic sources accelerate charged particles to relativistic speeds through mechanisms like shock acceleration and magnetic reconnection, after which the particles are injected into the interstellar medium. Consequently, the observed Galactic cosmic-ray population results from a combination of two fundamental plasma astrophysical processes: the acceleration of cosmic-ray particles and their subsequent transport through a turbulent medium. The interplay between acceleration and diffusion shapes the primary cosmic-ray spectrum. The propagation of cosmic rays is described by the following diffusion equation \citep{2007ARNPS..57..285S,2017JCAP...02..015E}:

\begin{equation}
    \begin{aligned}
        \frac{\partial \psi(\vec{r}, p, t)}{\partial t}= & Q(\vec{r}, p, t)+\vec{\nabla} \cdot\left(D_{x x} \vec{\nabla} \psi-\vec{V}_c \psi\right) \\
        & +\frac{\partial}{\partial p}\left[p^2 D_{p p} \frac{\partial}{\partial p} \frac{\psi}{p^2}\right] \\
        & -\frac{\partial}{\partial p}\left[\dot{p} \psi-\frac{p}{3}\left(\vec{\nabla} \cdot \vec{V}_c\right) \psi\right]-\frac{\psi}{\tau_f}-\frac{\psi}{\tau_r}
        \end{aligned}
        \label{eq1}
\end{equation}
where $ \psi(\vec{r}, p, t)$ represents the CR density per unit of total particle momentum $\rm p$ at position $ \vec{r}$, $ Q(\vec{r}, p, t)$ describes the source term, $\rm D_{x x}$ denotes the spatial diffusion coefficient, $ \vec{V}_c$ is the convection velocity and $\rm \tau_f$ and $\rm \tau_r$ are the timescales for loss by fragmentation and radioactive decay, respectively. The propagation of charged particles in the Galaxy is typically confined to a cylindrical region centered on the Galactic Center, where $\rm z$ represents the vertical height above the Galactic plane. 
The key physical quantity describing the diffusion behavior of cosmic rays, the diffusion coefficient $\rm D_{xx}$, is a function of particle rigidity (R), while $\rm D_{pp}$ represents the diffusion coefficient in momentum space. To simplify calculations, we assume that cosmic ray diffusion is uniform and isotropic. The specific expression for the diffusion coefficient D(R) is given by
\begin{equation}
D_{x x}(\mathcal{R})=D_0 \beta^\eta\left(\frac{\mathcal{R}}{\mathcal{R}_0}\right)^\delta
\label{eq2}
\end{equation}
Here, $\delta$ is the index of the diffusion coefficient, $\rm R_0$ is the reference rigidity, and $\rm D_0$ is the standardization factor of the diffusion coefficient.
The source term $ Q(\vec{r}, p, t)$ of eq.~\ref{eq1} describes the spatial distribution and injection power spectrum of cosmic ray sources, characterizing both the source locations and the momentum spectrum of injected particles. In this work, SNRs are considered as the primary sources of cosmic rays below the knee region (referred to as the background component). Their spatial distribution in cylindrical coordinates can be described as \citep{1996A&AS..120C.437C}:
\begin{equation}
f(r, z) \propto\left(r / r_{\odot}\right)^{1.69} \exp \left[-3.33\left(r-r_{\odot}\right) / r_{\odot}\right] \exp \left(-|z| / z_s\right)
\end{equation}
where $\rm r_{\odot}=8.5~kpc$ and $\rm z_s=0.2~kpc$. The accelerated cosmic rays escape from the source regions and enter the interstellar medium. Taking into account the nonlinear effects in the shock acceleration process of supernova remnants \citep{1999JETP...89..391B,2002APh....16..429B}, which soften the energy spectrum at low energies and harden it at high energies, a broken power-law injection spectrum is adopted in the form:
\begin{equation}
Q(\mathcal{R})=Q_0 \times \begin{cases}\left(\frac{\mathcal{R}}{\mathcal{R}_{b r}}\right)^{-\nu_1}, & \mathcal{R}<\mathcal{R}_{b r} \\ \left(\frac{\mathcal{R}}{\mathcal{R}_{b r}}\right)^{-\nu_2} \exp \left(-\mathcal{R} / \mathcal{R}_{c u t}\right), & \mathcal{R} \geq \mathcal{R}_{b r}\end{cases}
\end{equation}
where $\rm R=pc/Ze $ is the rigidity, $Q_0$ is the normalization factor for all nuclei, $R_{br}$ is the break rigidity, and $\nu_1$	and $\nu_2$ are the spectral indices before and after the break, respectively. It should be noted that an exponential cutoff factor $\rm exp(-R/R_{cut})$ is introduced here to describe the cutoff feature of the background cosmic-ray spectrum, with $\rm R_{cut}$ being the cutoff rigidity. This is based on long-term observational results suggesting that there may be an upper limit to the acceleration of cosmic-ray particles by SNRs \citep{2013MNRAS.431..415B,2018MNRAS.479.4470M}. In this work, cosmic rays at higher energies are assumed to originate from a single microquasar or a microquasar-like binary system.

\begin{figure}[t]
    \centering
    \includegraphics[width=0.95\linewidth]{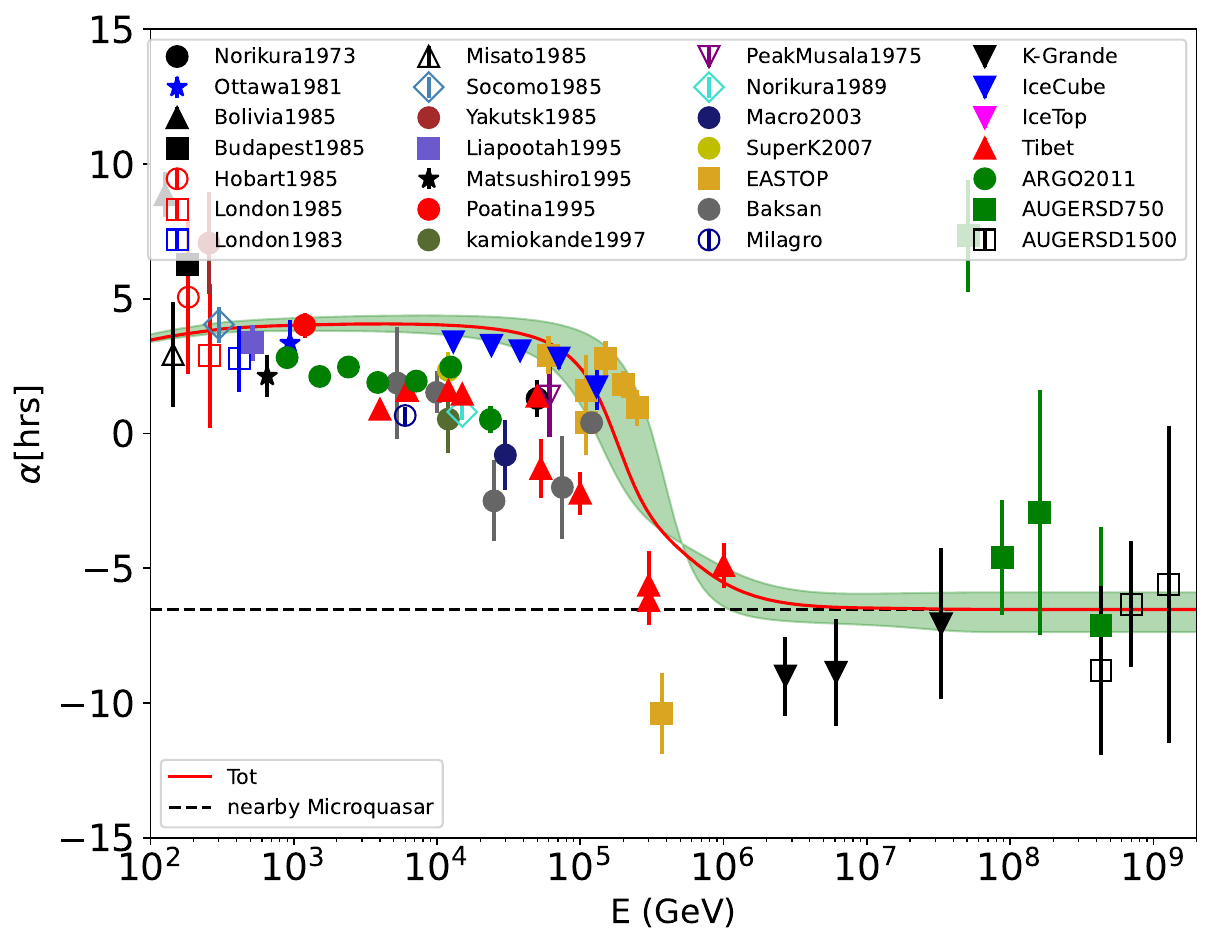}
    \caption{The energy dependence of the phases of the dipole anisotropies when adding all of the major elements together. The data points are taken from Norikura \citep{1973ICRC....2.1058S}, Ottawa \citep{1981ICRC...10..246B}, Bolivia \citep{1985P&SS...33.1069S}, Budapest \citep{1985P&SS...33.1069S}, Hobart \citep{1985P&SS...33.1069S}, London \citep{1985P&SS...33.1069S}, Misato \citep{1985P&SS...33.1069S}, Socomo \citep{1985P&SS...33.1069S}, Yakutsk \citep{1985P&SS...33.1069S}, Liapootah \citep{1995ICRC....4..639M}, Matsushiro \citep{1995ICRC....4..648M}, Poatina \citep{1995ICRC....4..635F}, kamiokande1 \citep{1997PhRvD..56...23M}, PeakMusala \citep{1975ICRC....2..586G}, Norikura \citep{1989NCimC..12..695N}, Macro \citep{2003PhRvD..67d2002A}, SuperK \citep{2007PhRvD..75f2003G}, EAS-TOP \citep{1995ICRC....2..800A,1996ApJ...470..501A,2009ApJ...692L.130A}, Baksan \citep{1987ICRC....2...22A}, Milagro \citep{2009ApJ...698.2121A}, K-Grande \citep{KASCADEAniso2015}, IceCube \citep{2010ApJ...718L.194A,2012ApJ...746...33A,2025ApJ...981..182A}, IceTop \citep{2013ApJ...765...55A}, AS-$\gamma$ \citep{2005ApJ...626L..29A,2017ApJ...836..153A,2015ICRC...34..355A}, ARGO \citep{2018ApJ...861...93B}, AUGER \citep{2024ApJ...976...48A}.}
    \label{alpha}
    \end{figure}
\begin{figure}[t]
    \centering
    \includegraphics[width=0.95\linewidth]{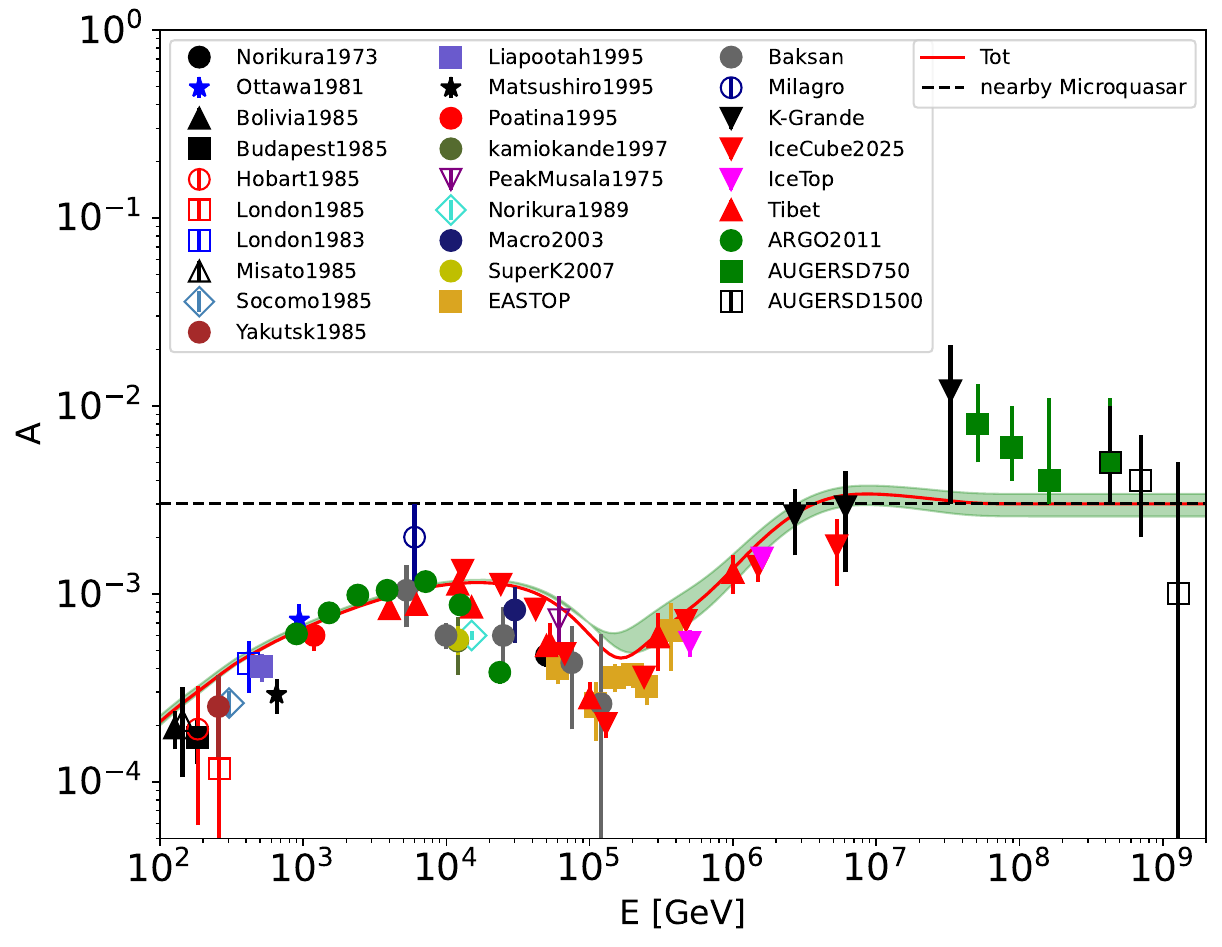}
    \caption{The same as Fig.\ref{alpha} but for the amplitudes.}
    \label{A}
    \end{figure}
\begin{figure}[t]
    \centering
    \includegraphics[width=0.95\linewidth]{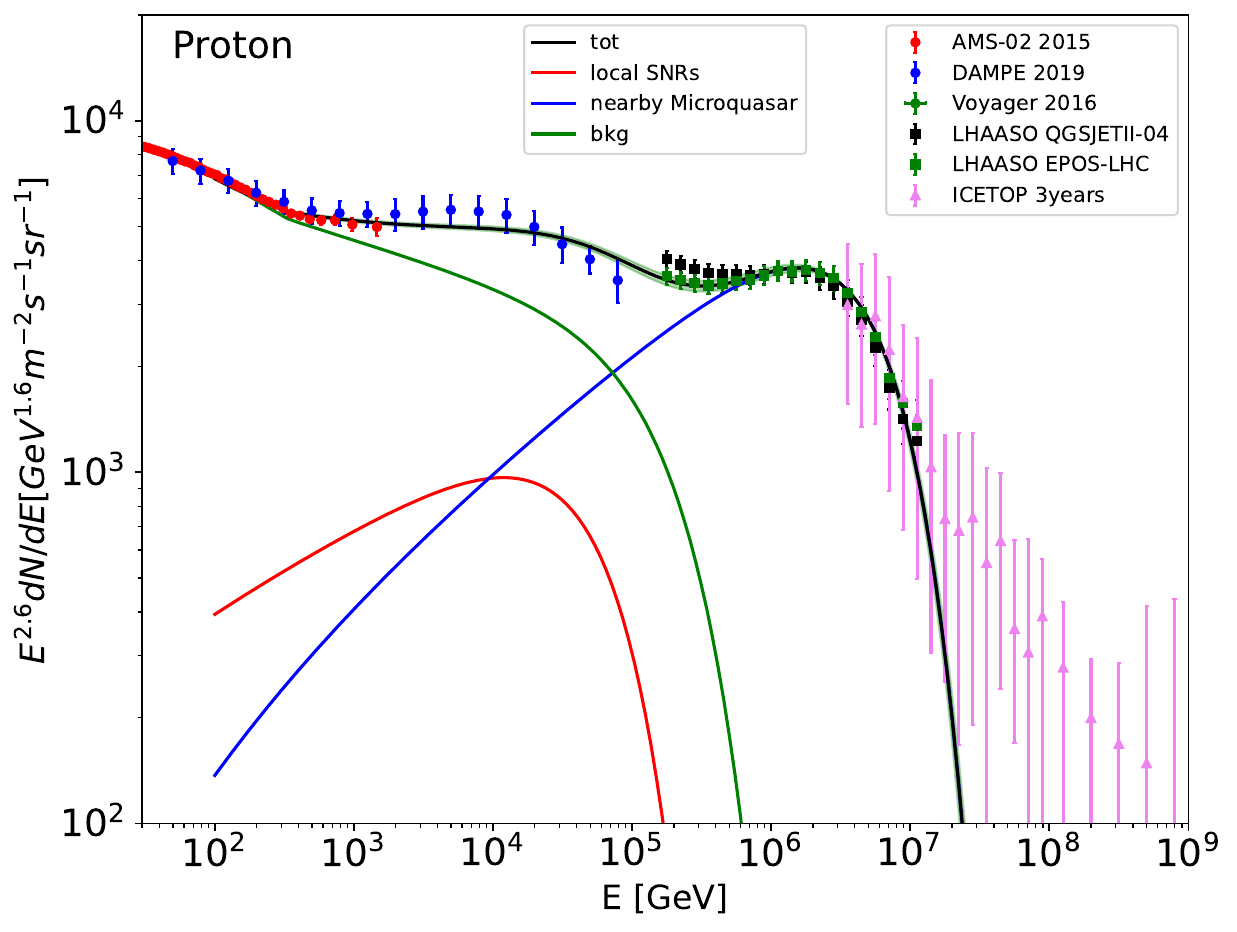}
    \caption{Same as figure \ref{proton1}, but the blue line shows the spectrum from the macroquasar located at the ideal position.}
    \label{proton2}
    \end{figure}
\begin{figure}[t]
    \centering
    \includegraphics[width=0.95\linewidth]{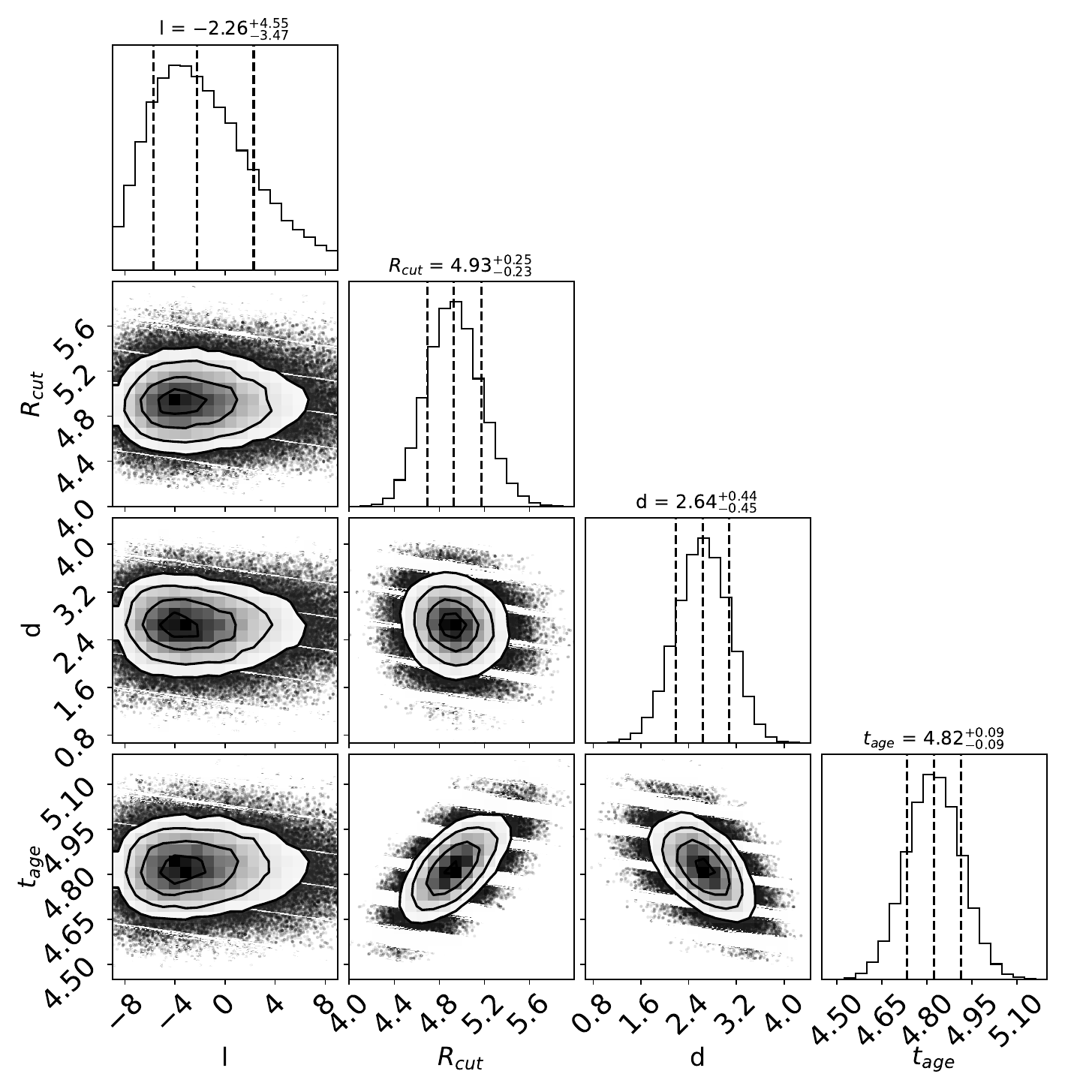}
    \caption{The one-dimensional probability distributions (diagonal) and two-dimensional credible regions for the model parameters. The parameter l is scaled by a factor of 0.1, and its allowed range is $[-180\degree,180\degree]$. The unit of break rigidity, distance and age is $\rm PV$, $\rm kpc$ and $\rm 10^6$ years, respectively.}
    \label{corner}
    \end{figure}
\subsection{Nearby Sources}
Recent LHAASO observations indicate that the microquasars SS 433, V4641 Sgr, GRS 1915+105, MAXI J1820+070, and Cygnus X‑1 are capable of accelerating cosmic rays to energies beyond the PeV range. It is therefore essential to investigate their contribution to the PeV cosmic-ray spectrum. In this work, we assume that the flux and spectral structure of cosmic rays in the knee region are predominantly governed by a single nearby source, while the background component produced by Galactic SNRs exhibits a cutoff in the sub‑PeV energy range. To reasonably constrain the properties of the source dominating the knee-region spectrum, we model the injection spectrum of this single source as an instantaneous injection, described by the following expression:
\begin{equation}
q(\mathcal{R})=q_0 \cdot\left(\frac{\mathcal{R}}{\mathcal{R}_0}\right)^{-\beta} \cdot e^{-\mathcal{R} / \mathcal{R}_{\text {cut},loc}}
\label{eq5}
\end{equation}
where $q_0$ is the normalization factor, $\rm R_0$ is the reference rigidity, and $\rm R_{cut,loc}$ represents the maximum rigidity of particles accelerated by the source. For simplicity, the diffusion of these single sources is assumed to be spherically symmetric and uniform. Using the diffusion coefficient given in Equation \ref{eq2}, the solution for the propagated spectrum from a single source can be written as:
\begin{equation}
\phi(r, t, \mathcal{R})=\frac{q(\mathcal{R})}{(\sqrt{2 \pi} \sigma)^3} \exp \left(-\frac{r^2}{2 \sigma^2}\right)
\label{eq6}
\end{equation}
Here, $\rm r$ is the distance from the source to Earth, $\rm t$ is the particle injection time, and $\sigma(\mathcal{R}, t)=\sqrt{2 D(\mathcal{R}) t}$ represents the effective diffusion radius of cosmic-ray particles within time $\rm t$.

It should be noted that although this study primarily focuses on cosmic rays in the knee region, observed spectral features—such as the hardening around 200 GeV for proton spectrum \citep{2015PhRvL.114q1103A,2015PhRvL.115u1101A}, the softening around 14 TeV \citep{2018JETPL.108....5A,2019SciA....5.3793A}, and the anisotropy phase pointing toward the anti‑Galactic center in this energy range—indicate the necessity of including a nearby cosmic-ray source component. Unlike the single microquasar mentioned above, the nearby source considered here belongs to the same category as the background-producing sources, namely SNRs. Consequently, although Equations \ref{eq5} and \ref{eq6} are also used to describe the contribution from the nearby source ( single SNR), the corresponding parameters differ significantly due to the substantial physical differences between microquasars and SNRs.

\subsection{Anisotropy of Cosmic Rays}
The background cosmic rays exhibit a cutoff in the sub-PeV energy range, while Galactic cosmic rays at PeV and higher energies are predominantly contributed by a single microquasar or a microquasar-like binary system. Consequently, the phase and amplitude of anisotropy in the PeV and higher energies should also be dominated by this source. By constraining the observed anisotropy phase and amplitude above PeV, it is possible to infer the location of this PeV source.

A key aspect of our study is to identify the essential parameters of the source that dominates cosmic rays at PeV and higher energies. To achieve this, we must first ensure a good fit to the observational data of cosmic rays below the sub-PeV range. We adopt the conventional cosmic-ray propagation model to compute the background component and additionally introduce a nearby SNR source in the anti-Galactic center direction to jointly account for the observational features of cosmic rays below 100 TeV. The introduction of a nearby SNR serves two purposes: it reasonably explains the hardening of the cosmic-ray proton spectrum around 200 GeV and its softening around 14 TeV, while also naturally accounting for the dip structure observed in the anisotropy amplitude from about 200 GeV up to the knee region, as well as the reversal of the anisotropy phase from the anti-Galactic center direction to the Galactic center direction at around 100 GeV.

\begin{figure}[t]
    \centering
    \includegraphics[width=0.95\linewidth]{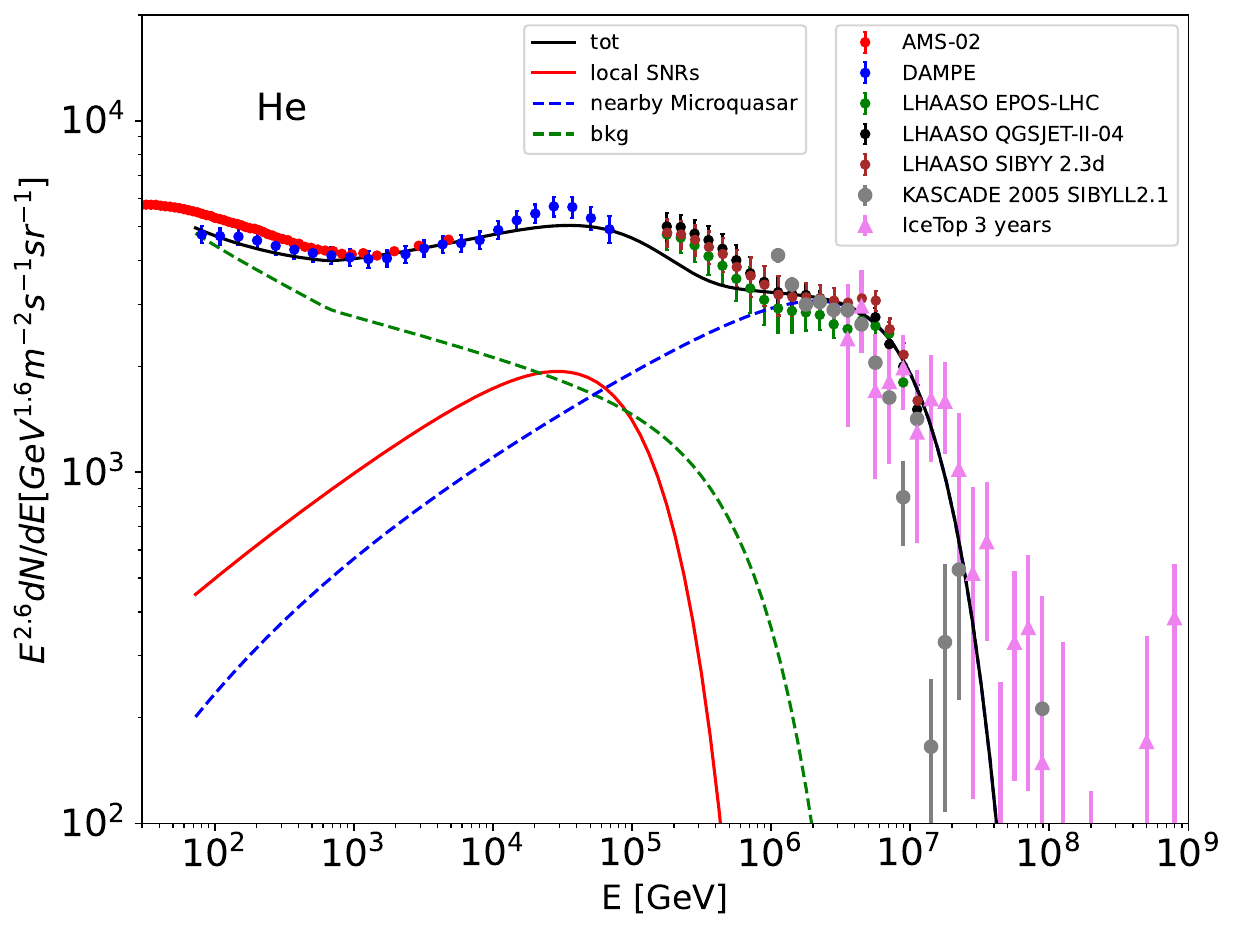}
    \caption{The CR  helium spectrum is calculated using the CR propagation model and compared with observational data from AMS-02 \citep{2015PhRvL.114q1103A}, DAMPE \citep{2019SciA....5.3793A}, LHAASO \citep{lhaasocollaboration2025precisemeasurementcosmicray}, KASCADE \citep{2014A&A...569A..32M} and ICETOP \citep{2019PhRvD.100h2002A}. }
    \label{he}
    \end{figure}
\begin{figure}[t]
    \centering
    \includegraphics[width=0.95\linewidth]{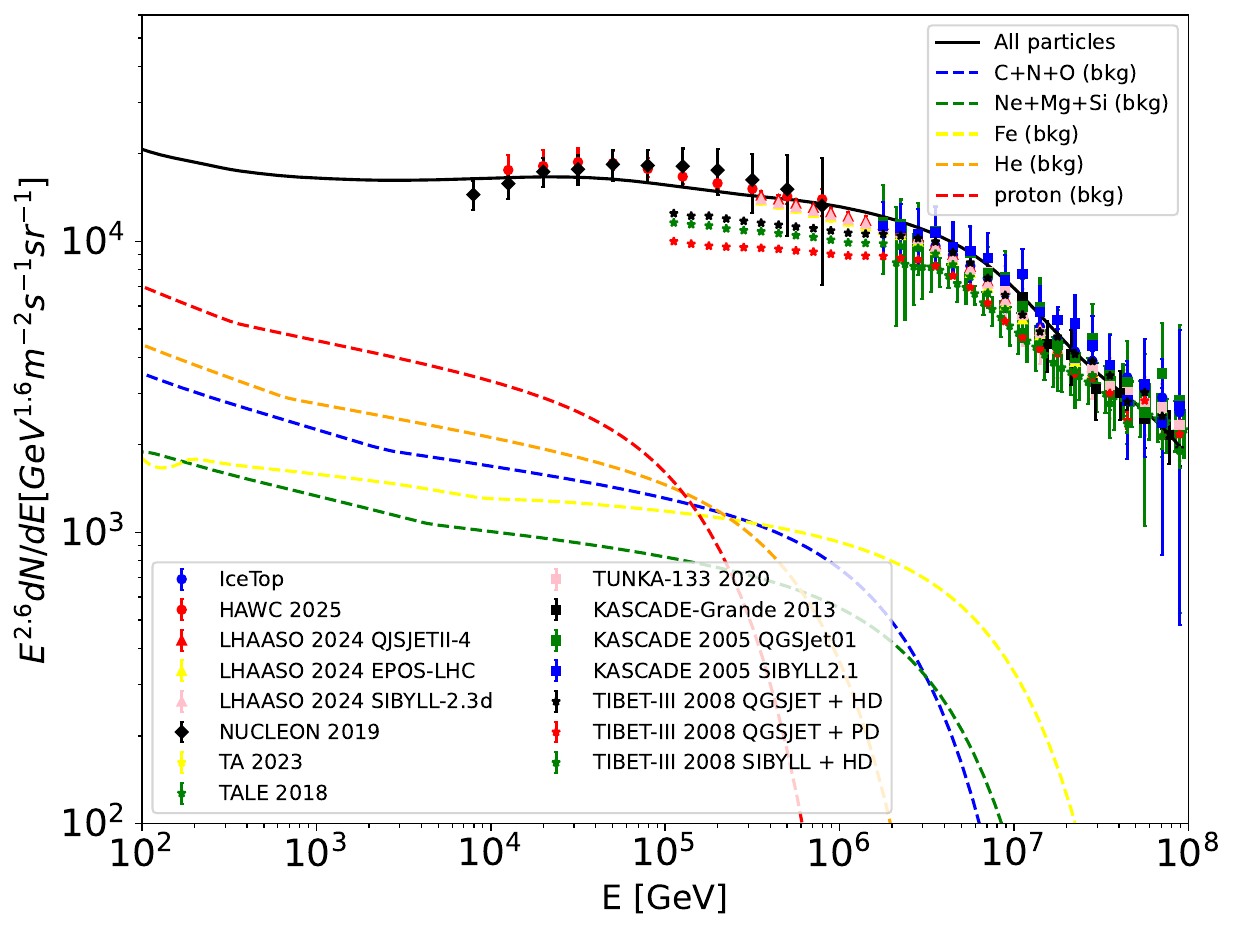}
    \caption{Model prediction of the all-particle spectra, compared with the observed data from Tibet-III \citep{2008ApJ...678.1165A}, TALE \citep{2018ApJ...865...74A}, TA \citep{2023APh...15102864A}, NUCLEON \citep{2019AdSpR..64.2546G}, KASCADE \citep{2024JCAP...05..125K}, HAWC \citep{2017PhRvD..96l2001A}, KASCADE-Grande \citep{2013APh....47...54A}, TUNKA-133 \citep{2020APh...11702406B} and LHAASO \citep{LHAASOspectrum2024prl}.}
    \label{allparticle}
    \end{figure}
\begin{figure*}[t]
    \centering
    \includegraphics[width=0.95\linewidth]{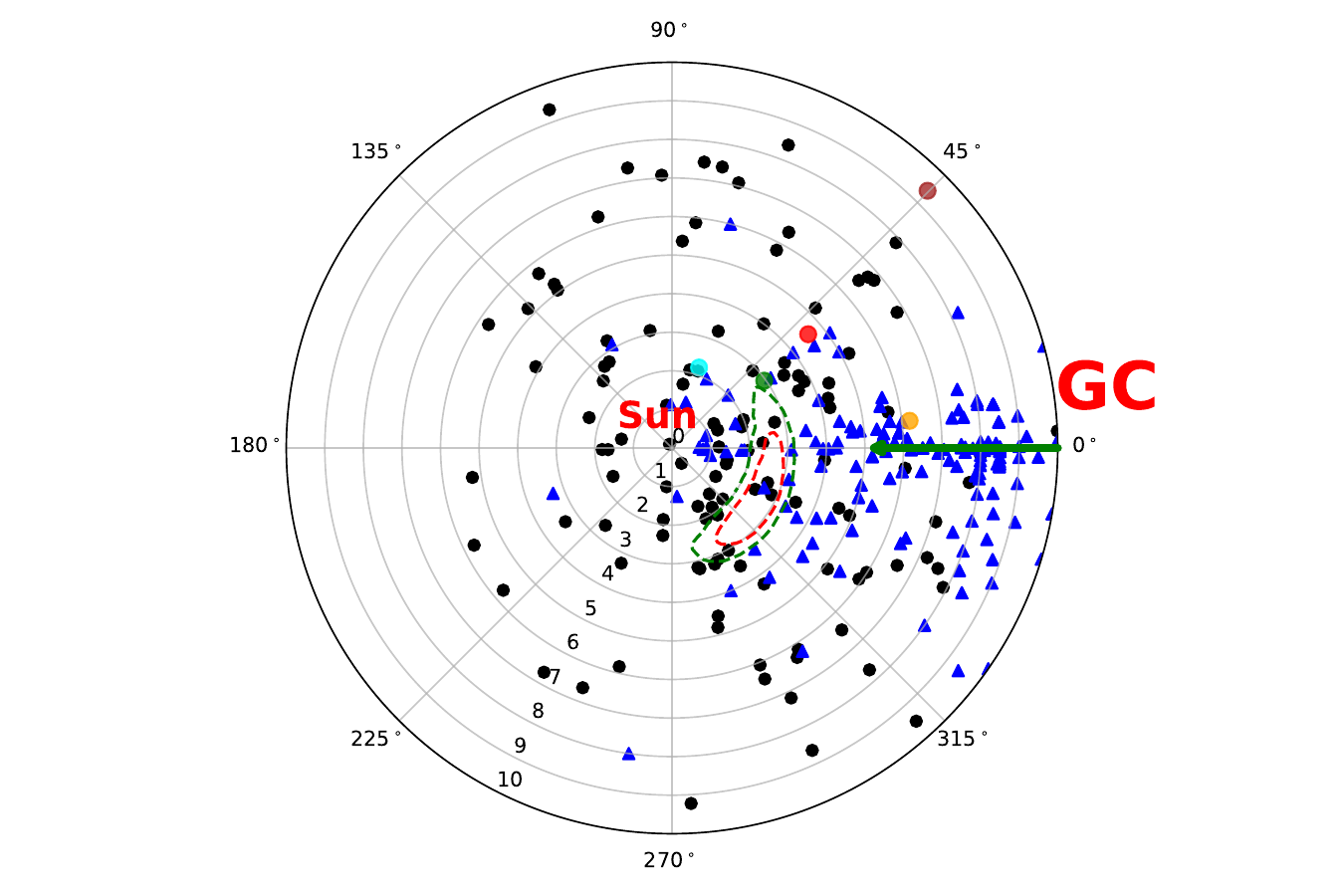}
    \caption{2D contour plot of the source distribution. The black and blue data points represent high-mass X-ray binaries (HMXBs) and low-mass X-ray binaries (LMXBs) on the Galactic disk ($\rm |b|<5\degree$) \cite{2023A&A...677A.134N,2023A&A...675A.199A}, respectively. Marked in the plot are SS 433 (red), V4641 Sgr (orange), GRS 1915+105 (brown), MAXI J1820+070 (green), and Cygnus X‑1 (cyan).  The dashed lines are 1 $\sigma$ (red) and 2 $\sigma$ (green) contours jointly constrained by observed data, respectively.
    The green arrow indicates the direction toward the Galactic Center.}
    \label{polar}
    \end{figure*}

\section{Results and Discussion} \label{sec:result}
We use the Galprop package to solve Equation \ref{eq1} and obtain the background cosmic-ray component \citep{1998ApJ...509..212S}. Then, within the same propagation environment, we compute the gamma-ray radiation from the point sources SS 433, V4641 Sgr, GRS 1915+105 and MAXI J1820+070, while evaluating their contribution to cosmic rays.

To calculate the contribution of propagated cosmic-ray protons near Earth from SS 433, V4641 Sgr, GRS 1915+105, and MAXI J1820+070, we employ the method described in Section \ref{sec:method}, using the background cosmic-ray spectrum and the secondary-to-primary ratio (B/C) to constrain the propagation model parameters. The halo half height and Galactic radius are set to $\rm H = 4.5~kpc$ and $\rm R=20~kpc$, respectively. Other relevant model parameters are determined by fitting low-energy B/C and cosmic-ray proton observational data, with the specific values listed in Table \ref{tab1}. Figure \ref{BC} show the theoretical predictions for the cosmic-ray B/C ratio at Earth, alongside the corresponding observational data.

To evaluate the contribution of SS 433, V4641 Sgr, GRS 1915+105, and MAXI J1820+070 to the cosmic-ray spectrum, the particle spectra injected by these sources into the interstellar medium and their ages must be constrained. Here we use the gamma-ray data observed by LHAASO to limit the respective injection spectra, and then employ the propagation model described above for the background cosmic rays to simulate their contribution at Earth. It should be noted that the diffusion timescale of particles from these sources is an important physical quantity. If we consider the injection from these four sources as continuous, their ages serve as a key reference for quantifying the diffusion timescale. For binary systems such as microquasars, their ages are difficult to determine precisely through observation; 1 Myr is commonly adopted as a typical value \citep{2021ApJ...910..149O}. In this work, except for SS 433, the other four binary systems are assigned this typical age. The age of SS 433 is believed to lie in the range of 10–100 kyr \citep{Cao_2025}; here we adopt the same value used by the LHAASO collaboration, 20 kyr, with its luminosity and proton spectral index constrained by observational data (see Table \ref{tab2}).

Figure \ref{gamma} displays the model-calculated gamma-ray spectra of SS 433, V4641 Sgr, GRS 1915+105, and MAXI J1820+070 together with the LHAASO observational data. As shown, the LHAASO data are well reproduced. Under the constraints of the gamma-ray observations, the cosmic-ray proton spectra near Earth produced by these microquasars are calculated, with the results presented in Figure \ref{proton1}. The calculations reveal that the contributions from SS 433, V4641 Sgr, GRS 1915+105, and MAXI J1820+070 are more than two orders of magnitude lower than the observational data.

A supplementary discussion on the contribution from Cygnus X‑1 is warranted. As one of the few sources detected by LHAASO with gamma-ray emission above 100 TeV, Cygnus X‑1 is located at a distance of about 2.2 kpc from Earth, making it the nearest microquasar among the five PeVatrons currently observed by LHAASO. However, because the diffusion coefficient inside the Cygnus bubble is more than an order of magnitude smaller than that in the general interstellar medium, previous studies have shown that cosmic-ray particles accelerated within the Cygnus bubble—even with the injection spectrum constrained by gamma-ray observations of the bubble—contribute negligibly to the cosmic-ray spectrum observed at Earth \citep{2024ApJ...974..276N}.

Returning to the central idea of this work, namely that the knee-region spectrum is dominated by a single microquasar or microquasar-like binary system relatively close to Earth. As a point source, its anisotropy amplitude is given by $\rm A=\frac{3 r}{2 c t}$, and its anisotropy phase depends almost exclusively on its directional location, with little coupling to other parameters. We therefore assume that such a nearby microquasar lies in the Galactic plane ($\rm |b|<5\degree$) and jointly fit the observed proton spectrum, anisotropy phase and amplitude data using its latitude, break rigidity, distance, and age as free parameters by the MCMC sampling method. Note that because the injection of background cosmic rays is cut off at 300 TV and the cosmic-ray phase begins to shift from the Galactic center direction to the anti-Galactic center direction around 2 EeV, we jointly fit the anisotropy data observed by KASCADE-Grande and AUGER as well as proton spectrum detected by LHAASO. The fitting results are shown in Figures \ref{alpha}, \ref{A}, \ref{proton2} and \ref{corner}. Additionally, Figures \ref{he} and \ref{allparticle} illustrate the agreement between the observed helium, all-particle spectra and the theoretical expectations. Overall, the results indicate that a microquasar or binary system located in the anti-Galactic center direction ($l \sim  337.4^{+44.5}_{-34.7} $) degree, approximately distance $2.64^{+0.44}_{-0.45}$ kpc from Earth, and with an age of about $4.82^{+0.09}_{-0.09} \times10^6$ years, can satisfactorily reproduce the observed proton, helium, all-particle spectra and anisotropy features.

\cite{2023A&A...677A.134N} and \cite{2023A&A...675A.199A} recently  published an updated catalog of Galactic low-mass X-ray binaries (LMXBs) and high-mass X-ray binaries (HMXBs). The catalog, based on homogeneous X-ray detections from Swift/XRT, Chandra, and XMM-Newton, contains 172 HMXBs and 360 LMXBs. These observational data with $|b|<5\degree$ are presented in Figure \ref{polar}, where the red contours marks the ideal region; 
the binary systems located in this region, jointly constrained by the MCMC within $1~\sigma$ (see Figure \ref{corner}), include IGR J16195-4945, IGR 16374-5043, IGR 16465-4507, IGR 17544-2619, and XTE J1650-500. One or several of these sources may contribute significantly to the knee region.
We hope that longer-term observations by detectors capable of observing gamma rays above 100 TeV, such as LHAASO, will help to further identify such candidates.

\section{Conclusion and Summary} \label{sec:conclusion}
This work proposes that Galactic cosmic rays primarily consist of two components: cosmic rays below the knee region are mainly accelerated by Galactic SNRs, and this component exhibits a cutoff in the sub-PeV range due to acceleration limits; whereas the flux of cosmic rays at and above the knee region is dominated by a single microquasar or microquasar-like binary system located near Earth. Under this framework, to investigate the properties of the single source responsible for the knee region, we used gamma-ray spectral observations as constraints to separately calculate the contributions of the microquasars recently observed by LHAASO—SS 433, V4641 Sgr, GRS 1915+105 and MAXI J1820+070—to the knee-region cosmic rays. The results indicate that the cosmic rays produced by these five sources contribute negligibly to the cosmic-ray spectrum at Earth. 

To identify the source that aligns with our hypothesis and can dominate the observed cosmic-ray spectrum and structure in the knee region, we performed a joint fit to the observed cosmic-ray proton spectrum, anisotropy phase, and amplitude, treating the source's latitude, break rigidity, distance from Earth, and age as free parameters. The fitting results reveal that among currently observed binary systems, five sources (IGR J16195-4945, IGR 16374-5043, IGR 16465-4507, IGR 17544-2619, and XTE J1650-500) show spatially good agreement with model predictions. Our findings provide several candidates as a useful reference for future observations, and we hope that longer-term data accumulation by LHAASO will yield more direct evidence in the future.

\section*{Acknowledgements}
This work is supported in China by National Key R$\&$D program of China under the grant 2024YFA1611402,
and supported by the National Natural Science Foundation of China (12333006, 12275279, 12375103).

\bibliographystyle{aasjournal}
\bibliography{ref_v2}{}
\end{document}